\documentclass[11pt]{article}
\pdfoutput=1
\usepackage{amsfonts,amsmath}
\usepackage{amssymb}
\usepackage{fancyhdr}
\usepackage{slashed}
\usepackage{graphicx}
\usepackage{subfigure}
\usepackage{color}

\usepackage{hyperref}
\usepackage[utf8]{inputenc}
\usepackage[titletoc]{appendix}

\def\hybrid{\topmargin -20pt    \oddsidemargin 0pt
        \headheight 0pt \headsep 0pt
        \textwidth 6.25in       
        \textheight 9.25in       
        \marginparwidth .875in
        \parskip 5pt plus 1pt   \jot = 1.5ex}

\hybrid

\def\baselinestretch{1.2}

                                                                                                                                                                                                                                                                                            \catcode`\@=11

\def\marginnote#1{}
\newcount\hour
\newcount\minute
\newtoks\amorpm
\hour=\time\divide\hour by60
\minute=\time{\multiply\hour by60 \global\advance\minute by-\hour}
\edef\standardtime{{\ifnum\hour<12 \global\amorpm={am}%
        \else\global\amorpm={pm}\advance\hour by-12 \fi
        \ifnum\hour=0 \hour=12 \fi
        \number\hour:\ifnum\minute<10 0\fi\number\minute\the\amorpm}}
\edef\militarytime{\number\hour:\ifnum\minute<10 0\fi\number\minute}

\def\draftlabel#1{{\@bsphack\if@filesw {\let\thepage\relax
   \xdef\@gtempa{\write\@auxout{\string
      \newlabel{#1}{{\@currentlabel}{\thepage}}}}}\@gtempa
   \if@nobreak \ifvmode\nobreak\fi\fi\fi\@esphack}
        \gdef\@eqnlabel{#1}}
\def\@eqnlabel{}
\def\@vacuum{}
\def\draftmarginnote#1{\marginpar{\raggedright\scriptsize\tt#1}}

\def\draft{\oddsidemargin -.5truein
        \def\@oddfoot{\sl preliminary draft \hfil
        \rm\thepage\hfil\sl\today\quad\militarytime}
        \let\@evenfoot\@oddfoot \overfullrule 3pt
        \let\label=\draftlabel
        \let\marginnote=\draftmarginnote
   \def\@eqnnum{(\theequation)\rlap{\kern\marginparsep\tt\@eqnlabel}%
\global\let\@eqnlabel\@vacuum}  }

\def\preprint{\twocolumn\sloppy\flushbottom\parindent 2em
        \leftmargini 2em\leftmarginv .5em\leftmarginvi .5em
        \oddsidemargin -.5in    \evensidemargin -.5in
        \columnsep .4in \footheight 0pt
        \textwidth 10.in        \topmargin  -.4in
        \headheight 12pt \topskip .4in
        \textheight 6.9in \footskip 0pt
        \def\@oddhead{\thepage\hfil\addtocounter{page}{1}\thepage}
        \let\@evenhead\@oddhead \def\@oddfoot{} \def\@evenfoot{} }

\def\numberbysection{\@addtoreset{equation}{section}
        \def\theequation{\thesection.\arabic{equation}}}

\def\underline#1{\relax\ifmmode\@@underline#1\else
        $\@@underline{\hbox{#1}}$\relax\fi}

\def\titlepage{\@restonecolfalse\if@twocolumn\@restonecoltrue\onecolumn
     \else \newpage \fi \thispagestyle{empty}\c@page\z@
        \def\thefootnote{\fnsymbol{footnote}} }

\def\endtitlepage{\if@restonecol\twocolumn \else \newpage \fi
        \def\thefootnote{\arabic{footnote}}
        \setcounter{footnote}{0}}  

\catcode`@=12
\relax

\def\figcap{\section*{Figure Captions\markboth
        {FIGURECAPTIONS}{FIGURECAPTIONS}}\list
        {Figure \arabic{enumi}:\hfill}{\settowidth\labelwidth{Figure
999:}
        \leftmargin\labelwidth
        \advance\leftmargin\labelsep\usecounter{enumi}}}
 \relax
\def\tablecap{\section*{Table Captions\markboth
        {TABLECAPTIONS}{TABLECAPTIONS}}\list
        {Table \arabic{enumi}:\hfill}{\settowidth\labelwidth{Table
999:}
        \leftmargin\labelwidth
        \advance\leftmargin\labelsep\usecounter{enumi}}}
 \relax
\def\reflist{\section*{References\markboth
        {REFLIST}{REFLIST}}\list
        {[\arabic{enumi}]\hfill}{\settowidth\labelwidth{[999]}
        \leftmargin\labelwidth
        \advance\leftmargin\labelsep\usecounter{enumi}}}
 \relax
\makeatletter
\newcounter{pubctr}
\def\publist{\@ifnextchar[{\@publist}{\@@publist}}
\def\@publist[#1]{\list
        {[\arabic{pubctr}]\hfill}{\settowidth\labelwidth{[999]}
        \leftmargin\labelwidth
        \advance\leftmargin\labelsep
        \@nmbrlisttrue\def\@listctr{pubctr}
        \setcounter{pubctr}{#1}\addtocounter{pubctr}{-1}}}
\def\@@publist{\list
        {[\arabic{pubctr}]\hfill}{\settowidth\labelwidth{[999]}
        \leftmargin\labelwidth
        \advance\leftmargin\labelsep
        \@nmbrlisttrue\def\@listctr{pubctr}}}
 \relax
\makeatother
\newskip\humongous \humongous=0pt plus 1000pt minus 1000pt

\newif\ifdtup

\relax

\def\be{\begin{equation}}
\def\ee{\end{equation}}
\def\ba{\begin{eqnarray}}
\def\ea{\end{eqnarray}}

\def\k{\kappa}
\def\r{\rho}
\def\a{\alpha}

\def\b{\beta}

\def\g{\gamma}

\def\d{\delta}

\def\e{\epsilon}

\def\th{\theta}

\def\m{\mu}
\def\n{\nu}

\def\l{\lambda}

\def\s{\sigma}

\def\cN{{\cal N}}

\def\cL{{\cal L}}

 \def\cK{{\cal K}} \def\cL{{\cal L}}
 \def\cN{{\cal N}} \def\cO{{\cal O}}

\newcommand{\prt}[1]{{\left( {#1} \right)}}

\def\no{\noindent}

\def\IR{\relax{\rm I\kern-.18em R}}

\def\pp{\partial}

\newcommand{\ff}{\frac}

\def\IR{\relax{\rm I\kern-.18em R}}
\def\IL{\relax{\rm I\kern-.18em L}}

\def\inv{^{\raise.15ex\hbox{${\scriptscriptstyle -}$}\kern-.05em 1}}

\def\cL{{\cal L}}

\def\bea{\begin{eqnarray}}
\def\eea{\end{eqnarray}}
\newcommand{\eq}[1]{(\ref{#1})}
\def\nn{\nonumber}

\newcommand{\la}[1]{\label{#1}}

\def\a{\alpha}      
\def\b{\beta}       
\def\g{\gamma}    
\def\d{\delta}    
\def\e{\epsilon}

\def\k{\kappa}
\def\l{\lambda} 
\def\m{\mu} \def\n{\nu}

\def\r{\rho}
\def\s{\sigma}  
\def\t{\tau}
\def\th{\theta}

\def \dt {\dot{t}}

\definecolor{markcolor2}{rgb}{1,0,0}

\definecolor{markcolor3}{rgb}{0,1,0}

\begin{document}

\begin{titlepage}

\begin{center}


~
\vskip 1 cm

{\large
\bf Chaotic Motion near Black Hole and Cosmological Horizons}

\vskip 0.4in
{\bf Dimitrios Giataganas$^{1,2}$  }
\vskip 0.2in
{\em
  ${}^1$  Department of Physics, National Sun Yat-sen University,  \\
Kaohsiung 80424, Taiwan
\vskip .1in
 ${}^2$  Center for Theoretical and Computational Physics,   \\
Kaohsiung 80424, Taiwan
\\ \vskip .15in
{\tt dimitrios.giataganas@mail.nsysu.edu.tw}
}

\vskip .4in
\end{center}

\vskip 1.4in

\centerline{\bf Abstract}
It is known that certain types of particle motion near black hole horizons are chaotic while it has been proposed the existence of a universal bound for their Lyapunov exponent. We discuss the relation between chaos and inaffinity in presence of black hole and cosmological horizons. We argue that although a relation between the Lyapunov exponent and the generalized surface gravity appears naturally, in general there is no reason for the Lyapunov exponent of classical trajectories to be bounded in generic spacetimes with  horizon. Moreover,  we show that the de Sitter spacetime and cosmological horizons act as a nest of chaos in holography and we find that the Lyapunov exponent of the trajectories is related to the inaffinity in the same way for both cosmological and black hole horizons. This suggests that there is no distinction by the Lyapunov exponent between maximal chaos of black hole and cosmological horizons.
\no
\end{titlepage}
\vfill
\eject


\noindent


\def\baselinestretch{1.2}
\baselineskip 19 pt
\noindent


\setcounter{equation}{0}

\section{Introduction}

In gravity the geodesic motion convey important features of the background spacetimes and has been extensively studied.   For example, the gas in the accretion disk approaches the black hole through nearly circular orbits losing at the same time its angular momentum. Through this sequence the gas reaches the innermost stable circular orbit (ISCO), the closest point to the horizon of stable circular motion. Beyond this point the gas follows a radial accelerated motion to the black hole.  Therefore the ISCO effectively coincides with the inner edge of accretion disk and as a result the properties of the circular time-like motion can be used to estimate the spin of observed black holes \cite{Zhang:1997dy,Narayan:2005ie}. Moreover, the instabilities of circular orbits have been found to be fundamentally related to the merging of the black holes \cite{Pretorius:2007jn} providing indirect information of this dynamical process. Additionally, the instability of null geodesics is also related to the gravitational collapse of stars \cite{1968thorn} and the quasinormal modes of the black holes \cite{Kokkotas:1999bd,Nollert:1999ji}. These are only some of the examples where it is evident that the study of geodesics reveals significant properties of the black holes and collapsing stars. This connection is not surprising, in contrary it is expected and it is a usual way to extract information from the physical systems. The geodesics act as probes in the system and their properties are related to the properties of the heavy gravitational objects sourcing strong curvature effects on the space-time. It is natural for the properties of gravitational field to be reflected on the geodesic motion. Furthermore, the fact that geodesics may develop instabilities and  chaos it is also expected due to the non-linear nature of gravitational fields in general relativity.

Recently, there has been an intense effort to quantitatively understand the properties of chaos around black holes, in quantum systems and in the context of holography. It is very likely that there exist a systematic way to relate the appearance and the properties of chaos in these very different systems  making use of the underlying link of the fundamental laws of complexity. Some of the earlier works on the relation between the dual operators and chaos, include the study of the chaotic sting motion in the context of holography, which  reveals the nonintegrable nature of the dual field theories, for example \cite{Zayas:2010fs,Giataganas:2013dha,Giataganas:2017guj}. More recently, it has been formulated an equivalence between the 2-dim $\sigma$-model spectrum expanded on a nontrivial massive vacuum and the motion of a classical particle in a non-trivial potential. In particular, the presence of chaos of a classical particle, is equivalent to the non-factorization of the $S$-matrix in the dual quantum 2-dim theory \cite{Giataganas:2019xdj}. These developments consisted of some initial studies on the fundamental significance of the classical chaos and its potential relevance in quantum field theories through holography.

In a parallel direction there have been studies of the chaos properties in quantum systems. These studies were initiated by the seminal conjectured bound on the leading Lyapunov exponent $\l$ of the out-of-time-ordered correlators (OTOC) in thermal quantum field theories, that is equal to $2 \pi T/\hbar$ \cite{Maldacena:2015waa}, where $T$ is the temperature. The proposal provided a stepping stone to many further related developments in the gauge/gravity duality and black hole physics. In a more classical context, it has motivated the study of the particle motion near black holes under suitable external potential which turned out to realise, at least in certain cases, a similar bound. The  Lyapunov exponent for certain trajectories is bounded by the surface gravity at the horizon of the spherically symmetric static black hole \cite{Hashimoto:2016dfz}. This relation has been studied further and challenged in several setups \cite{Dalui:2018qqv,Zhao:2018wkl,Lei:2020clg,Cubrovic:2019qee,Colangelo:2021kmn}. In another relevant direction it has been also proposed an interesting correspondence between the operator growth in chaotic theories, the complexity and the radial momenta of the particles falling in the AdS black hole, for example in \cite{Susskind:2018tei,Magan:2018nmu,Brown:2018kvn,Ageev:2018msv,Barbon:2019tuq}.

In this letter we continue the studies of the properties of chaotic trajectories in the near horizon regime and elaborate further on the relation between the Lyapunov exponent and the inaffinity or equivalently the surface gravity. We argue that in the general the dependence of the Lyapunov exponent to the generalized surface gravity we define appears naturally. We find that the bounds on the Lyapunov coefficient are not necessarily satisfied in generic black hole spacetimes even after imposing certain energy conditions. We present such examples to support the generic analysis and to show that the null energy conditions are not enough to constrain maximal chaos on the horizons. Moreover, we extend the current studies to include the cosmological horizons in the context of holography. The study on de Sitter is partly motivated by recent developments on OTOC and related studies in dS spacetimes \cite{Aalsma:2020aib,Chapman:2021eyy,Jacobson:2021nbb,Anninos:2018svg}. We find that, from the point of view of the classical trajectories, the properties of chaos around holographic de Sitter horizons present similarities with the chaos generated by the black hole horizons. The Lyapunov exponent of the trajectories is related to the inaffinity in the same way irrespective the type of horizon, black hole or cosmological one. This implies that the properties of maximal chaos of quantum field theories in curved space-times resemble the ones of maximal chaos in thermal field theories in flat space-times although for non-maximal chaos there can be a distinction. This result can be thought as being the classical parallel with the OTOC computations in de Sitter space-times,  in the same way that the proposed bound on certain black holes \cite{Hashimoto:2016dfz} happens to match with \cite{Maldacena:2015waa}. The  OTOC of four conformally coupled scalar fields exhibits maximal chaos saturating the proposed chaos bound \cite{Aalsma:2020aib}. Our classical trajectories in the Anti-de Sitter space-times with de Sitter slicing saturate the classical bound and there is no distinction between maximal chaos in black holes and de Sitter space-times by the leading Lyapunov exponent.

\section{Chaos Generation by Horizons}

Let us consider a $d+2$-dimensional gravitational theory with the generic characteristics of a homogeneous black hole given by
\be \la{metric}
ds^2=g_{tt}(r) dt^2+g_{rr}(r) dr^2+g_{ii}(r) dx_i^2+g_{rr}(r) dr^2~.
\ee
The background has a horizon $r_h$ where $g_{tt}(r_h)=0$, and is allowed to have a boundary where $g_{ii}(r_b)\rightarrow \infty$, which we take without loss of generality to be at infinity.

\subsection{The Inaffinity and the Effective Surface Gravity Function}

The geodesic equation of a particle in the  space-time \eq{metric} when parametrized by a parameter $\s$ that is not affine
is given by
\be
\frac{d^2 x^\mu}{d \s^2}+ \Gamma^\mu_{\nu\l} \frac{d x^\nu}{d \s} \frac{d x^\l}{d \s} = \k (\s) \frac{d x^\m}{d \s}~.
\ee
The inaffinity coefficient $\k$, parametrizes the failure of the parameter $\s$ to be affine parameter and can be found by the following integral
\be
\frac{d \t}{d\s}= e^{\int^{\s} d\bar{\t} \k(\bar{\t})}~,
\ee
where $\t$ is an affine parameter. This is the realization of the statement that the geodesic itself is not reparametrization invariant and retains its form only under certain affine transformations related to the proper time $\t$.
The inaffinity has a known physical significance, which becomes evident in the near horizon black hole regime where the lack of certain coordinates to be affine related to properties of the black holes and the chaotic motion of geodesics.

Let us consider a null normal vector $\xi$ ($\xi^\a \xi_\a$=0) on any null hypersurface $\cN$ which generates a family of null geodesics
\be\la{null}
\xi^\a \nabla_\a \xi^\beta =\k \xi^\b~,
\ee
where $\nabla$ is the covariant derivative. Affine parametrization of the geodesics would correspond to vanishing $\k$. Once we match the hypersurface $\cN$ with the Killing horizon, the vector field $\xi$ becomes a Killing one and we can not make an arbitrary coordinate transformation to reach to an affine parametrization. In particular, then we can only rescale $\xi$ by a constant. Thus, $\k$ is defined  uniquely as long as we choose asymptotically a normalization for $\xi$. Then  $\k$  acquires an additional  meaning  and can be identified to the surface gravity of the Killing horizon.

The surface gravity can be computed by the Killing vector, a  straightforward process for stationary black holes \cite{Blaupp}. The Killing horizon in this case is defined as
\be
\cK= \{y: S(y):=-\xi^\a(y) \xi_\a(y) =0 \}~,
\ee
where $\pp_\a S$ is normal to $\cK$ and therefore parallel to $\xi_\a$. The proportionality factor between $\pp_\a S$ and $\xi_\a$, is equal to the surface gravity as
\be
\nabla_\b\prt{-\xi^\a \xi_\a}=2 \k \xi_\b~,
\ee
where we have used equation \eq{null}.
We like to express $\k$ in terms of the black hole metric spacetime. We proceed by using the equivalent definition of the surface gravity that involves a manipulation based on the Frobenius integrability condition $\xi_{[\a}\nabla_\b \xi_{\g]}=0$, which is a direct consequence of the properties of the vector $\xi$. Note that here the brackets '${[]}$' denote a full antisymmetrization. Taking into account that $\xi$ is a Killing vector
\be\la{ak1}
\nabla_\a \xi_\b+\nabla_\b\xi_\a=0~,
\ee
the Frobenius condition remains with the three independent terms:
$\xi_{\a}\nabla_\b \xi_{\g}+\xi_{\g}\nabla_\a \xi_{\b}+\xi_{\b}\nabla_\g \xi_{\a}=0$. We contract it with the  with $\nabla^\a \xi^\b$ to generate  $\k$, then by using \eq{ak1} and \eq{null} we get
\be\la{surface}
\k^2=-\frac{1}{2}\prt{\nabla^\a \xi^\b}\Big(\nabla_\a \xi_\b \Big) ~.
\ee
This is the known formula of the surface gravity, encoding the rate of change of the null vectors and  it is a direct measure of the strength of the gravitational field.

Therefore the surface gravity is related to the acceleration, the measure of the gravitation force on a static observer at a surface $S(r)$ as seen by an observer at infinity. To realize it in a more convenient form, we compute the coordinate invariant norm of the acceleration  $\mathbf{\a}(r)^2:= \a^\mu \a_\mu $ and take into account the redshift factor $\mathbf{V}:=\prt{-V^\a V_\a}^{1/2}$ associated with the timelike Killing vector $V= \pp_t$.
The norm of the Killing vector for the metric \eq{metric} is $\mathbf{V}(r)^2= -g_{tt}$ and we normalize the 4-velocity $u^\mu$ of a static observer to have unit magnitude as $u^\mu= V^\mu/\mathbf{V}$. The unit vector of the static observer then reads $u^\mu=\prt{\prt{-g_{tt}}^{-1/2}, 0,\ldots,0}$. The acceleration of the trajectory is defined as $\a^\mu=u^\nu \nabla_\nu u^\mu$ and since we refer to the static observer it is non-zero. Then in the background \eq{metric} the proper acceleration is equal to
\be\la{acceler}
\mathbf{\a}(r)=\ff{ g_{tt}'(r)}{2 \sqrt{g_{rr}(r)} g_{tt}(r)}
\ee
which diverges at the horizon of the black hole reflecting the difficulty of the static observer to remain static in the near horizon regime.

Putting everything together, we define  for our purposes the useful function $\k_{eff}(r)$, such that $\lim_{r\rightarrow r_h}\k_{eff}=\k$, which we will refer to as the generalized or effective surface gravity. Using the acceleration of the fiducial observer \eq{acceler} at $r$ we get
\be\la{surfaceef}
\k_{eff}(r)= \mathbf{V}(r)\mathbf{\a}(r)=- \frac{g_{tt}'(r)}{2\sqrt{-g_{tt}(r) g_{rr}(r)}}~,
\ee
which can be thought as the surface gravity of a virtual sphere of radius $r$. When $\k_{eff}$ is computed at the horizon $r_h$ by the formula  \eq{surfaceef}, it coincides the well defined surface gravity $\k$ associated with the Killing horizon of the black hole
\be
\k= - \frac{g_{tt}'(r_h)}{2\sqrt{-g_{tt}(r_h) g_{rr}(r_h)}}~.
\ee
$\k$ is finite and is the acceleration of a static observer at the horizon as measured by an observer at infinity. It is also the inaffinity, the measure of the lack of affinity of the family of geodesics associated to the normal vector field $\xi$, or the failure of the corresponding coordinates to be affine at the horizon.

Assuming that we approach the horizon from infinity, or that there exist a boundary of the theory at $r=\infty$, $g_{tt}$ is a monotonically decreasing function and therefore the effective surface gravity is positive. On the other hand, the monotonicity of the $\k_{eff}$ depends on the properties of the space and does not always develop a maximum at the horizon. We are mostly interested in the near horizon regime, where we get
\be\la{kexp}
\k_{eff}(r)=\k+\ff{-g_{tt}'\prt{g_{rr} g_{tt}}'+2g_{rr} g_{tt} g_{tt}''}{4\prt{-g_{tt} g_{rr}}^{3/2}}\bigg|_{r_h}\prt{r-r_h}+\cO\prt{(r-r_h)^2}~.
\ee
The behavior of the effective function depends on the symmetries and the dimension that the gravitational background respects. For instance the term $g_{rr} g_{tt}$, can be a monotonically decreasing function (or constant) for a class of Lifshitz geometries that respect the Null Energy Conditions, therefore the first term in the fraction is negative. The term $2 g_{rr} g_{tt} g_{tt}''$ takes the opposite sign of the second derivative of the time metric element and it can be positive or negative.  Therefore the overall correction term is not guaranteed to have a certain sign for physical theories and as a result the monotonicity of $\k_{eff}$ can not be fixed in general as we approach the horizon. The redshifting factor in geometries with boundary at infinity plays a major role in this behavior.  Nevertheless, for the AdS geometries indeed $\k_{eff}$ develops a local maximum at the horizon in the near horizon regime  equal to the surface gravity for $d\ge 2$. For other theories, for instance with Lifshitz scaling symmetry, $\k_{eff}$'s monotonicity depends on the dimension and the scaling coefficient.
We will elaborate more on this in the following sections.

\subsection{Relating the Lyapunov Exponent to the Inaffinity} \la{sect:rely}

The relation of the surface gravity to the Lyapunov exponent can be understood by general relativity arguments at least for certain type of geodesics based on the analysis of the role of inaffinity presented in the previous section.  We consider an equilibrium surface for the trajectories at $r_0$ in the near horizon regime $r_h$. When the equilibrium is unstable  and exists chaos for radial trajectories, these are described by an exponential divergence of the form $r(t)\sim e^{\l t} $, where $\l$ is the Lyapunov exponent and has to depend on the gravitational and external potentials characterizing the motion.
On the other hand, as we have shown in the previous section, the (effective) surface gravity measures the strength of the acceleration in the near horizon regime as seen by an observer at infinity and  the failure to have an affine parametrization. Therefore, the acceleration of the radial trajectories and as a result the Lyapunov exponent should depend on the surface gravity.
A non-singular relation of the Lyapunov exponent to the surface gravity or inaffinity, implies a finite Lyapunov exponent at horizon and in the near horizon regime since the surface gravity is finite. For a certain class of theories which have a redshifted radial acceleration that is larger at the horizon  as seen by an observer at infinity,  the particle will experience the most intense chaos in the near horizon regime. For these cases an upper bound equal to the surface gravity of the Killing horizon
can be justified as it  has been proposed by \cite{Hashimoto:2016dfz} for spherically symmetric static black holes.

\subsection{Lyapunov Exponents and Linearized Equations of Motion}

The Lyapunov exponents can be introduced in a general framework for a particle motion with Lagrangian $\cL$ \cite{Cardoso:2008bp}. The equations of motion can be written as
\be
\ff{d y^i}{dt}=F_i(x)~,
\ee
where $F$ depends on the Lagrangian and its derivatives. A linerization reads
\be
\ff{d\d y^i}{dt}= \ff{\pp F_i }{\pp y^j} \d y^j(t)~,
\ee
with a solution described by
\be
\d y^i(t)=L_{ij} \d y^j(0)~,\qquad \dot{L}_{ij}(t)=\ff{\pp F_i }{\pp y^l} L_{lj}(t)~,
\ee
where $L_{ij}$ is the evolution matrix. The Lyapunov exponent $\l$ is the measure of the spatial divergence of the trajectory with respect to its initial point. From the eigenvalues of the evolution matrix $L$ after we allow the system to evolve for long enough time, $\l$ can be obtained by
\be
 \l=\lim_{t\rightarrow \infty} \ff{1}{t}\log \ff{L_{ii}(t)}{L_{ii}(0)}~.
\ee
Limiting the motion on two-dimensional phase space $y^i=(p_r , r)$, for example  of orbits of constant $r$, the expressions simplify to
\bea
\ff{\pp F_1}{\pp y^2} =  \ff{d}{dr}\prt{\ff{1}{\dt}\ff{\d \cL}{\d r}}~,\qquad
\ff{\pp F_2}{\pp y^1} = -\ff{1}{\dt g_{rr}}~
\eea
and the principal Lyapunov exponent simplifies to
\be\la{lyap1}
\l=\sqrt{\ff{\pp F_1}{\pp y^2} \cdot \ff{\pp F_2}{\pp y^1}}~,
\ee
which can be computed directly from the equations of motion.

\subsection{Lyapunov Exponents and Geodesic Stability}

Let us be more precise and consider a particle with mass $m$ moving in an external potential $V$ in the generic spacetime \eq{metric}. The motion of the particle is given by the Lagrangian
\be
\cL=- m\sqrt{-g_{tt}}\sqrt{1+ \frac{g_{rr}}{g_{tt}} \dot{r}^2+\frac{g_{ii}}{g_{tt}} \dot{x}_i^2}-m V(r,x_i)~.
\ee
Provided that we work for  slow motion in the non-relativistic limit where the time-derivative terms are subleading to the rest of the terms, we rewrite the action as
\be
\frac{\cL}{m}= - \sqrt{-g_{tt}}\prt{ \frac{1}{2} \frac{g_{rr}}{g_{tt}} \dot{r}^2+\frac{1}{2} \frac{g_{ii}}{g_{tt}} \dot{x}_i^2}- V_{eff}(r)~,
\ee
where the effective potential is defined as
\be
V_{eff}(r) :=   V(r)+\sqrt{-g_{tt}}~.
\ee
In order to have an equilibrium point and a particle that does not fall in the black hole we need a negative competing external potential $V$ against the gravitational potential $V_{gr}=\sqrt{-g_{tt}}$. For an unstable equilibrium point $r_0$, it is enough to have $V'_{eff}=0$ and $V''_{eff}\le0$. This implies that $V'(r)$, should generate a repulsive external force counterbalances the gravitational force of the horizon.
Let us consider a  saddle point $r_0$ near the horizon. This is given by solving the algebraic equation
\be\la{stable1}
V'(r_0)=\frac{g_{tt}'(r_0)}{2 \sqrt{-g_{tt}(r_0)}}~.
\ee
In order to study the stability let us expand the potential around the saddle point $r_0$
\bea\la{effe}
&& L_{eff}=-V_{eff}(r_0)+\frac{g_{rr}(r_0)}{2\sqrt{-g_{tt}(r_0)}}\prt{ r'^2+\frac{g_{ii}(r_0)}{g_{rr}(r_0)} x_i'^2}- \frac{1}{2} V_{eff}''(r_0) \prt{r-r_0}^2~,\\\la{veff}
&& V_{eff}''(r_0) :=  V''(r_0)-\frac{g_{tt}''(r_0)}{2 \sqrt{-g_{tt}(r_0)}}-\frac{g_{tt}'(r_0)^2}{4 (-g_{tt}(r_0))^{3/2}}~.
\eea
Without loss of generality we can obtain the properties of the system by localizing the motion on the spatial coordinated $x_i$ and focusing on the radial direction. The derivation of the equation of motion is straightforward
\be
r''(t)+\ff{\sqrt{-g_{tt}(r_0)}}{g_{rr}(r_0)}V_{eff}''(r_0)\prt{r-r_0}=0~,
\ee
and has the analytical solution
\be\la{solr1}
 r(t)= r_0 + c_1 e^{\lambda t}+ c_2 e^{-\lambda t}~.
\ee
The particle dynamics are determined by
\be\la{lveff2}
\l^2= -\ff{\sqrt{-g_{tt}(r_0)}}{g_{rr}(r_0)} V_{eff}''\prt{r_0}~,
\ee
where $V_{eff}$ is given by \eq{veff}. The particle motion depends on the redshifted concavity (or convexity) of the effective potential scaled by the inverse radial metric element. Concave effective potential implies unstable saddle points for which $\lambda^2 >0$ and a solution \eq{solr1} exhibits an exponentially growing behavior.  Once the coupling to the other coordinates $x_i$ will be restored,   the $d$-dimensional  $\prt{r(t),x_i(t)}$ motion will develop chaos detectable for example in Poincare section.

\subsection{The Inaffinity in Chaos}\la{section:lyap}

The Lyapunov exponent \eq{lveff2} depends on the redshifting factor and the derivative of the time metric element.  We can use \eq{surfaceef} to express it with respect to effective surface gravity as
\be\la{lamda2}
\l^2=\k_{eff}^2+\ff{g_{tt}''(r_0)}{2 g_{rr}(r_0)  }\prt{1-\ff{2 \sqrt{-g_{tt}(r_0)}}{g_{tt}''(r_0)}V''(r_0)}~,
\ee
which justifies the Lyapunov dependence on $\k_{eff}$ as expected by the discussion of subsection \ref{sect:rely}.
To examine the near-horizon regime, we assume that we have an appropriate potential $V_{eff}$ that allows in this regime a saddle point. In the near horizon  \eq{lamda2} becomes
\bea\nn
&&\l^2=\k^2 + \ff{1}{2 g_{rr}}\prt{g_{tt}''-2 \sqrt{-g_{tt}}V''}\Big|_{r=r_h}+
\\\nn
&&\prt{\k_{eff}^{(1)}-\ff{g_{rr}'\prt{ g_{tt}''-2\sqrt{-g_{tt}} V''}}{2 g_{rr}^2}+\ff{g_{tt}' V''}{2 g_{rr}\sqrt{-g_{tt}}}+\ff{g_{tt}^{(3)}-2\sqrt{-g_{tt}} V^{(3)}}{2 g_{rr}}}\Bigg|_{r=r_h}\prt{r_0-r_h}+\l^2{}^{(2)}\prt{r_0-r_h}^2~
\eea
where the upper bracketed indices denote the near-horizon expansion terms and $\k_{eff}^{(1)}$ can be read from \eq{kexp}.  Notice that taking the limit to the horizon of the above expression leads to several  vanishing terms and that the near horizon limit of the potential dependent terms should be taken with caution, since at this limit the derivative of the potential at the saddle point behaves as \eq{stable1}
\be\la{stable2}
V'\sim  - \ff{\sqrt{-g_{tt}^{(1)}}}{ 2 \sqrt{r_0-r_h}}~.
\ee
The near horizon expression can be rewritten as
 \bea\la{lexp2}
\l^2=\k^2 + \ff{1}{4}\prt{G_{rr}^{(1)}g_{tt}^{(2)}-G_{rr}^{(2)}g_{tt}^{(1)}}\prt{r_0-r_h}-G_{rr}^{(1)}\sqrt{-g_{tt}^{(1)}} V''(r_h) \prt{r_0-r_h}^{3/2}+\cO((r_0-r_h)^2)~,
\eea
where $G_{rr}:=1/g_{rr}$ and the upper indices denote the finite component terms of the expansion, e.g. $g_{tt}(r)= g_{tt}^{(1)} \prt{r-r_h}+ g_{tt}^{(2)} \prt{r-r_h}^2 + \dots$. The $V''(r_h)$ has not been expanded in components and could potentially alter the order of its contribution $(r_0-r_h)$ in $\l^2$, depending on how it scales. For example, it can contribute first order terms, 
using \eq{stable2} in the external potential to absorb the near singularity  rewriting the potential dependent term of \eq{lexp2} as $2 \k^2 \tilde{V}'' /\tilde{V}' \prt{r_0-r_h}$ where $\tilde{V''}$ denotes the derivative of the regular part of the potential at the horizon. This possible substitution does not affect our following argument of the existence of no upper bound in Lyapunov.

The Lyapunov exponent is related to the effective surface gravity as
\bea\la{lexp3}
\l^2=\k_{eff}^2 + G_{rr}^{(1)}g_{tt}^{(2)}\prt{r_0-r_h}-G_{rr}^{(1)}\sqrt{-g_{tt}^{(1)}} V''(r_h) \prt{r_0-r_h}^{3/2}+\cO((r_0-r_h)^2)~,
\eea
where the $\k_{eff}^2$ includes its $(r_0-r_h)$ terms \eq{kexp}. We notice that there is no constrain a-priori that the right hand side of the above expression becomes maximal at the horizon. In fact  the $V$-independent terms are not monotonic functions, as we have already discussed \eq{kexp}. This expression already hints that the universality of the bound is limited only to certain class of theories. Moreover, since in principle there are no conditions to fully constrain the potential dependent term, these terms can not be solely responsible on setting a general bound. Nevertheless, it would be interesting to study how the application of additional energy conditions on given gravity actions coupled to fields, affect the chaos and its maximization.

We apply our analysis in the context of holography and in particular in theories with broken symmetry. It is well known that classical string solutions corresponding to dual operators of non-integrable theories develop chaos even in the vacuum. However, chaos is very rarely observed at zero temperature for operators that correspond to point-like string solutions \cite{Giataganas:2013dha}. At finite temperature our analysis suggests the existence of chaos in geodesics. Let us apply the analysis of Lifshitz black holes with
\be
g_{tt}=-r^{2z} f(r)~,\qquad g_{rr}=\ff{1}{r^2 f(r)}~,\qquad f(r)=1-\prt{\ff{r_h}{r}}^{d+z} ~,
\ee
where $d$ is the number of dimensions and $z$ is the dynamical critical exponent. This type of black holes lead to
\be
\k_{eff}=\ff{\prt{d+z} r_h^{z}}{2}-\ff{1}{2} \prt{r_h^{z-1}\prt{d-2z} \prt{d+z}}\prt{r_0-r_h}~,
\ee
where the first term correspond to the surface gravity. The function is decreasing as we move away of the horizon for $d\ge 2 z$ or $d\ge -z$, for the positive and negative values of z respectively. The Lyapunov exponent reads
\be
\l^2=\ff{\prt{d+z}^2 r_h^{2z}}{4}-\ff{1}{2} \prt{r_h^{2z-1}\prt{d+z}^2\prt{z-1}} \prt{r_0-r_h}-r_h^{z+\ff{1}{2}} \prt{d+z}^{3/2} V''(r_h) \prt{r_0-r_h}^{\ff{3}{2}}~.
\ee
Focusing on the independent terms of the external potential, we would get a maximum Lyapunov exponent bounded by the surface gravity for $z\ge 1$. This could be for example the case of linear external potentials. The satisfaction of the Lyapunov bound for $z\ge 1$ and its violation for the rest of the values, matches the Null Energy Conditions (NEC): $T_{\m\n} N^\m N^\n\ge0 \Leftrightarrow R_r^r-R_0^0\ge0 \Leftrightarrow z\ge 1$, where $T_{\m\n}$ is the energy momentum tensor contracted with the null vectors $N^\m$ and $R_\m^\n$ is the Ricci tensor. It takes the form
\be
\l^2_{Lif}=\k^2-\ff{2\k^2 r_h^{-1}}{d}\prt{R_r^r-R_0^0} \prt{r_0-r_h}-r_h^{z+\ff{1}{2}} \prt{d+z}^{3/2} V''(r_h) \prt{r_0-r_h}^{\ff{3}{2}}~.
\ee
The situation is  more involved for black holes with hyperscaling violation. The NEC depend on two parameters and their relation to the Lyapunov bound is not as straightforward. The metric of the isotropic hyperscaling black holes is
\be
g_{tt}=-r^{2z-\ff{2 \th}{d}} f(r)~,\qquad g_{rr}=\ff{1}{r^{2+\ff{2 \th}{d}} f(r)}~,\qquad f(r)=1-\prt{\ff{r_h}{r}}^{d+z-\th}~,
\ee
where $\th$ is the hyperscaling violation exponent. This type of black holes leads to
\be
\k_{eff}=\ff{\prt{d+z-\th} r_h^{z}}{2}-\ff{1}{2d}\prt{d\prt{d-\th-2z}+2 \th} \prt{d+z-\th} r_h^{z-1}\prt{r_0-r_h}~,
\ee
resulting to a decreasing effective surface gravity away of the horizon only in certain regime of $(\th,z)$. The Lyapunov exponent reads
\bea\nn
\l^2&=&\ff{\prt{d+z-\th}^2 r_h^{2z}}{4}-\ff{1}{2d} \prt{r_h^{2z-1}\prt{d+z-\th}^2\prt{d\prt{z-1}-2\th} }\prt{r_0-r_h}\\
&&-r_h^{z+\ff{1}{2}+\ff{\th}{d}} \prt{d+z-\th}^{3/2} V''(r_h) \prt{r_0-r_h}^{\ff{3}{2}}~.
\eea
The V-independent part of the Lyapunov exponent develops an upper bound at the surface gravity when $d\prt{z-1}-2\th\ge0$. The NEC for the hyperscaling violation theories give $R_r^r-R_0^0\ge0\Rightarrow d^{-1}\prt{d-\th}\prt{d\prt{z-1}-\th}\ge0$ and $R_x^x-R_0^0\ge0\Rightarrow\prt{z-1}\prt{d+z-\th}\ge0$. While there is an additional stability constrain $(d-\th)/z\ge0$ from the thermodynamics since the entropy scales as $S\sim T^\ff{d-\th}{z}$. To constrain further the parametric space on physical and stable theories we may also apply the condition that $\th\le d$ to guarantee that the entanglement entropy scales slower than the volume, consistent to a dual quantum field theory behavior \cite{Dong:2012se}. We can rewrite the Lyapunov exponent as
\be\la{lhysca2}
\l^2_{hysca}=\k^2-\ff{2 \k^2 r_h^{-1}}{d} \prt{\ff{d}{\prt{d-\th}}\prt{R_r^r-R_0^0}-\th}\prt{r_0-r_h}-r_h^{z+\ff{1}{2}+\ff{\th}{d}} \prt{d+z-\th}^{3/2} V''(r_h) \prt{r_0-r_h}^{\ff{3}{2}}~.
\ee
By applying all the physical constrains it is clear that the NEC do not necessarily lead to the maximal chaos on the horizon. Nevertheless,  we can numerically observe that for the most of the parametric space, when we restrict to physical theories, the V-independent terms maximize at the horizon. However, there is a narrow regime on the parametric space of corresponding to stable field theories that it still leads to violation of this behavior. It is caused by the different $\th$ coefficient in the condition $d\prt{z-1}-(2)\th\ge0$ which appears both in NEC and surface gravity.
Therefore, the energy and stability conditions although they tend to constrain the theory towards the parametric regime that the V-independent part develops an upper bound on the Lyapunov in \eq{lhysca2}, are not enough to fully guarantee it.

\section{Chaos Near Cosmological Horizons}

So far we have focused on the study of chaos in the near horizon regime of black hole horizons, being dual to thermal theories in flat space-times. An observer moving along a time-like geodesic in de Sitter space experiences a heat bath of particles at temperature $T_{dS}=H/2\pi$ when the theory is in the vacuum state, where $H$ is the cosmological horizon.  Although this thermal behavior resembles the one associated to the black hole horizons even in holography \cite{Chu:2016pea}, there are a couple of crucial differences:  $T_{ds}$ depends on the curvature on the spacetime and therefore the thermal effects are a direct consequence of the curvature, and the cosmological horizon is observer dependent.

Let us consider an AdS/dS metric by slicing the AdS space to have a de Sitter boundary using an appropriate embedding, which corresponds to a dual field theory that lives in a curved de Sitter space \cite{Chu:2016uwi}. The gravitational background can be also generated by solving the geometry in a perturbation expansion in Fefferman-Graham coordinates \cite{Marolf:2010tg} and in lower dimensions by appropriate transformations of the BTZ black hole \cite{Tetradis:2011jn}.
The theory is dual to the canonical choice of Bunch-Davies vacuum that reduces to the Minkowski space vacuum state in the limit $H\rightarrow 0$. We choose to view the state in the static patch where the thermal nature of the vacuum for the local observer is associated to the presence of the cosmological horizon.

The background reads
\bea
ds^2=\ff{1}{z^2}\prt{f(z)^2 ds^2_{dS}+dz^2}~,
\eea
where the boundary metric is the de Sitter
\be
ds^2_{dS}=\prt{-h(\r) dt^2+\ff{1}{1-H^2 \r^2} d\r^2+\r^2 d\Omega_{d-2}^2 }
\ee
and the corresponding functions are defined as
\be
f(z):=1-\ff{H^2 z^2}{4}~,\qquad h(\r):=1-H^2 \r^2~.
\ee
The  cosmological horizon is at $\r=1/H$ and the bulk horizon at $z_h=2/H$. We have a timelike Killing vector $\pp_t$ and the energy related to $T^{00}$ is conserved. The temperature of the dual field theory is given by $T_{dS}=H/\prt{2 \pi}$ and the stress tensor is regular on the boundary. Moreover, the stress tensor of the theory is not traceless since the theory expected to have an anomaly. In this theory we study whether the cosmological horizon generates chaos and how Lyapunov exponent behaves.

The horizon in this coordinate system bounds $\r$  from above $\r\le 1/H$, so a change the total sign of \eq{surfaceef} is needed. Here we bring the particle in the proximity of the cosmological horizon $\r=1/H$, while the external potential should create a repulsing force towards the origin $\r=0$. The test particle experiences a weak bulk gravitational potential, which is negligible since the unstable extremum $\r_0$ is placed in the near boundary regime $z=\e$ of the space and very close to the cosmological horizon. By taking this into consideration, the generic formalism of the subsection \ref{section:lyap} applies for the AdS sliced cosmological horizon.

The effective surface gravity in the near cosmological horizon regime and close to the boundary of the theory is computed by \eq{surfaceef} to give
\be
\k_{eff}=H^2 \r~.
\ee
The surface gravity at the cosmological horizon is then equal to $\k=H$, and it is related to the Hawking temperature with the known relation $T_{dS}=\k/\prt{2 \pi}$.  Its role is similar to the one of the static black hole:  $\k$ is the redshifted force to hold a test particle at the horizon. 

The Lyapunov exponent in the near horizon regime is computed by \eq{lexp3} to give
\be
\l^2=\k_{eff}^2\bigg|_{\r=\ff{1}{H}}~,
\ee
where we have taken the near boundary limit along $z$-direction and we have omitted the $V$-dependent terms. $\l$ depends on the effective surface gravity and the extra terms in the similar way with black hole horizons although the expansion away of the horizon is different. 

The Lyapunov exponent becomes maximal at the cosmological horizon and equal to the surface gravity of the horizon
\be
\l_{AdS/dS}=\k_{dS}~.
\ee
There is no distinction between maximal chaos in black holes and de Sitter spacetimes by the leading Lyapunov exponent.  This could have consequences on the chaotic nature of the expectation values of the dual operators and holographic observables, the maximal chaos is indistinguishable in black holes and the curved de Sitter. Cosmological horizons and black hole horizons are seen in the same way from the point of view of maximal chaotic motion, however their leading term expansion of the Lyapunov is differs.

\textbf{Acknowledgments:}
The research work of D.G. is supported by  Ministry of Science and Technology of Taiwan (MOST) by the Young Scholar Columbus Fellowship grant 110-2636-M-110-007.

\bibliographystyle{JHEP}

\end{document}